\documentclass[iop]{emulateapj}
\usepackage{color}

\usepackage{epstopdf}
\usepackage{amsmath}
\makeatletter
\newcommand\lsim{\mathrel{\rlap{\lower4pt\hbox{\hskip1pt$\sim$}}
\raise1pt\hbox{$<$}}}
\newcommand\gsim{\mathrel{\rlap{\lower4pt\hbox{\hskip1pt$\sim$}}
\raise1pt\hbox{$>$}}}

\makeatother
\shorttitle{The Dynamics  Kepler--56 Planetary System  }
\shortauthors{Li et al}
\makeatother
\begin{document}
\title{ The Dynamics of the Multi-planet System Orbiting Kepler-56 }

\author{Gongjie Li\altaffilmark{1}, Smadar Naoz\altaffilmark{1}\altaffilmark{$\dagger$}, Francesca Valsecchi\altaffilmark{2,3}, John Asher Johnson\altaffilmark{1}, Frederic A.~Rasio\altaffilmark{2,3}} 
\altaffiltext{1}{Harvard Smithsonian Center for Astrophysics, Institute for
  Theory and Computation, 60 Garden St., Cambridge, MA 02138}
\altaffiltext{2}{Center for Interdisciplinary Exploration and Research in Astrophysics (CIERA), Northwestern University, Evanston, IL 60208, USA}
\altaffiltext{3}{Department of Physics and Astronomy, Northwestern University, Evanston, IL 60208, USA}
\altaffiltext{$\dagger$}{Einstein Fellow}
\email{gli@cfa.harvard.edu;snaoz@cfa.harvard.edu}

\begin{abstract}
Kepler-56 is a multi-planet system containing two coplanar inner planets that are in orbits misaligned with respect to the spin axis of the host star, and an outer planet. Various mechanisms have been proposed to explain the broad distribution of spin-orbit angles among exoplanets, and these theories fall under two broad categories.
The first is based on dynamical interactions in a multi-body system, while the other assumes that disk migration is the driving mechanism in planetary configuration and that the star (or disk) is titled with respect to the planetary plane. Here we show that the large observed obliquity of Kepler 56 system is consistent with a dynamical origin. In addition, we use observations by \citet{Huber+13} to derive the obliquity's probability distribution function, thus improving the constrained lower limit. The outer planet may be the cause of the inner planets' large obliquities, and we give the probability distribution function of its inclination, which depends on the initial orbital configuration of the planetary system. We show that even in the presence of precise measurement of the true obliquity, one cannot distinguish the initial configurations. Finally we consider the fate of the system as the star continues to evolve beyond the main sequence, and we find that the obliquity of the system will not undergo major variations as the star climbs the red giant branch. We follow the evolution of the system and find that the innermost planet will be engulfed in $\sim 129$ Myr. Furthermore we put an upper limit of $\sim 155$ Myr for the engulfment of the second planet. This corresponds to $\sim 3\%$ of the current age of the star.
\end{abstract}

\section{Introduction}\label{intro}
Over the past few years, measurements of the sky-projected obliquity of exoplanets have found that large obliquities and even retrograde systems are common among hot Jupiters \citep[e.g.][]{FW09,Tri+10,Morton10, Moutou+11,Albrecht+12,Hebrard+13}.   Recently, \citet{Hirano+12}, \citet{Sanchis+12},  \citet{Albrecht+13}, \citet{Chaplin+13} and \citet{Van14} have measured the obliquity of six transiting multi-planet systems discovered by the NASA  {\it Kepler} mission, and found they all have low obliquities. However, \citet{Huber+13}, using asteroseismology, showed that large obliquities are not confined to Hot Jupiter systems. In fact Kepler-56 has two, low mass, inner planets whose orbit normal is tilted with respect to the stellar spin axis. 

Several mechanisms have been suggested to explain the formation of misaligned planets. These theories can be divided into two categories.
The first is based on tilting the orientation of an inner planet compared to the stellar spin axis. This category includes scattering and secular dynamical effects between a planet and a companion, or other planets in the system that can produce large obliquities \citep[e.g.,][]{Dan,Sourav+08,Nag+08,Naoz11,Naoz+12bin, Naoz13, WY11,Li+13,Li+14, ValsecchiRasio14b,ValsecchiRasio14a}. 
These mechanisms predict that an inner planet with a large obliquity has an outer perturber which is inclined with respect to the plane of the inner planet, the perturber can be either a stellar companion or a planet, or even multiple planets.
In the second category, planets move inward from their birthplaces beyond the snow line by migrating inward through the protoplanetary disk  \citep[e.g.][]{Lin+86,Mass+03}. 
Large obliquities can then be produced either by tilting the stellar spin axis with respect to the orbital angular momentum \citep[e.g.][]{Winn+10b,Lai+10,Rogers+12,Rogers+13IGW, Spalding14}, or by tilting the protoplanetary disk \citep{Bate10, Batygin12}. %
This second category of models predicts that the various planets in a system should lie roughly in the same plane since they were confined to the same flattened disk. 

Here we focus on the dynamical  mechanism that produced the large obliquities in the Kepler-56 planetary system. 
Most of the theoretical  studies investigating large obliquities focused on Hot Jupiters, mainly because these were observed to have large obliquities. The underlining physics of producing a misalignment  in the presence of a perturber is very similar. Thus,  such studies are relevant for investigating the Kepler-56 system (as we will show below). 

%
%
%

Kepler multiple systems are typically packed, small sized  \citep[$\sim 1-10$~R$_\oplus$ e.g.][]{Lissauer+11,Swift+13} and close--in \citep[$\sim 1-100$~d, e.g.,][]{Steffen+13} systems. At face value these configurations may indicate that dynamical and secular processes are suppressed, since these systems better resemble the theoretical outcome of planet migration in the protoplanetary disk, given their low mutual inclinations \citep{Lissauer+11, Fang12}. Therefore, a large obliquity in a multi-planet system may be used as a laboratory to test the two categories of models summarized above. In other words, since it seems that these planets form in a disk, a tilt of the protoplanetary disk or of the star, will cause the multiple planets to show the same obliquity. 
%

Kepler-56 is an evolved star at the base of the red giant branch in the HÑR diagram with $m_\star=1.32$~M$_{\odot}$   $R_\star= 4.23$~R$_\odot$ and an age of $3.5$~Gyr \citep{Huber+13}. Furthermore,  \citet{Huber+13} showed that the innermost planet ($m_b=0.07$~M$_J$, $R_b=0.65$~R$_J$,  hereafter planet ``b") has a period of $10.5$~d, and a period of $21.4$~d for the other planet ($m_c=0.57$~M$_J$, $R_c=0.92$~R$_J$,  hereafter planet ``c"). 
The mutual inclination between these two planets is measured to be $< 5^\circ$. Kepler-56 is an interesting system as it raises many questions regarding its formation and future evolution. Most importantly,  \citet{Huber+13}, measured the obliquity of the system using asteroseismology and placed a  lower limit on the true obliquity of the two inner planets of $\psi>37^\circ$. The dynamical analysis of \citet{Huber+13} favors the scattering and later torquing scenario.

Here we use Kepler-56's current observations to compute the probability distribution for its obliquity.
(\citet{Huber+13} reported observations already give enough information to calculate such distribution.) This enables us to also put strong constrains on the probability distribution of the outer planet's inclination with respect to the innermost two. Furthermore, we estimate that the two inner planets will be engulfed in $\sim 129$ Myr and $\lesssim 155$ Myr, respectively. The engulfment of the inner planets is consistent with the the deficit in short period planets around retired A stars \citep[e.g.][]{Johnson07, Sato08, Bowler10, Schlaufman13}.

The paper is structured as follows. We calculate the obliquity distribution function from observations, and show that the current observations give more information than just a lower limit (Section \ref{sec:psi}). We then  discuss the current obliquity precession as a function of the system initial conditions (Section \ref{sec:psiprec}) and show that combining the physical understanding and the observed distribution, we can infer the outer most planet orbital inclination with respect to the innermost two as a function of the initial configuration (Section \ref{sec:pofi}). We also calculate the orbit and obliquity future evolution as the star further ascends the giant branch  (Section \ref{sec:Sevol}). We finally offer our discussion (Section \ref{sec:dis}).

\begin{figure*}[hbt]
\begin{center}
\includegraphics[width=8cm,clip=true]{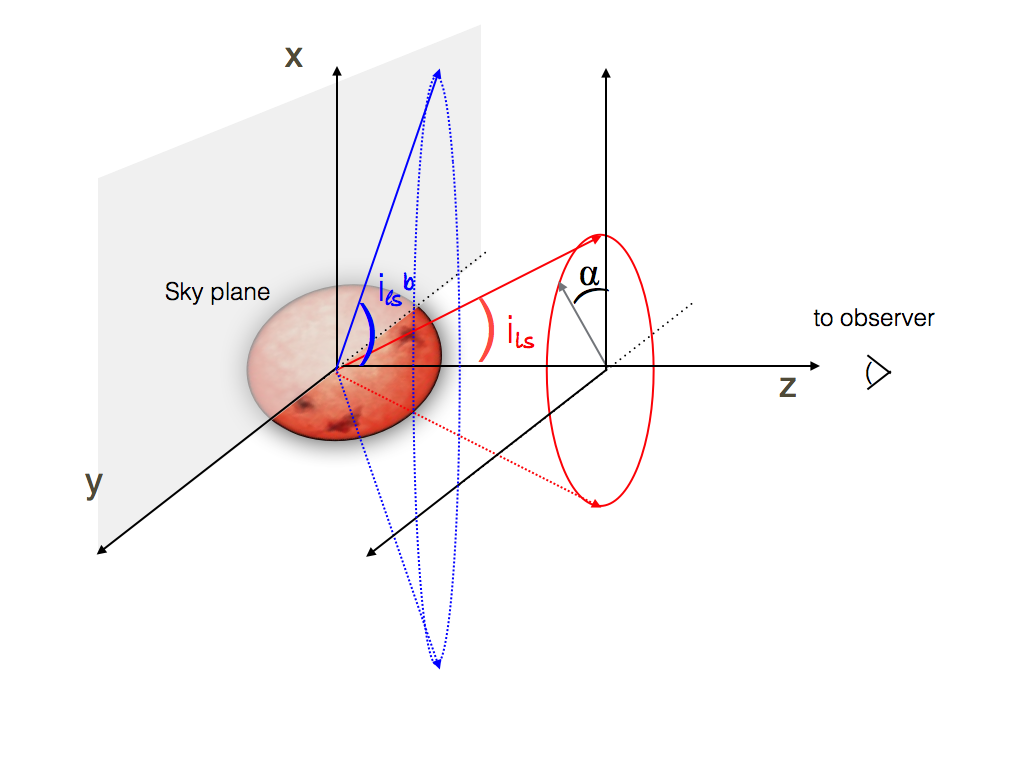}
\includegraphics[width=8cm,clip=true]{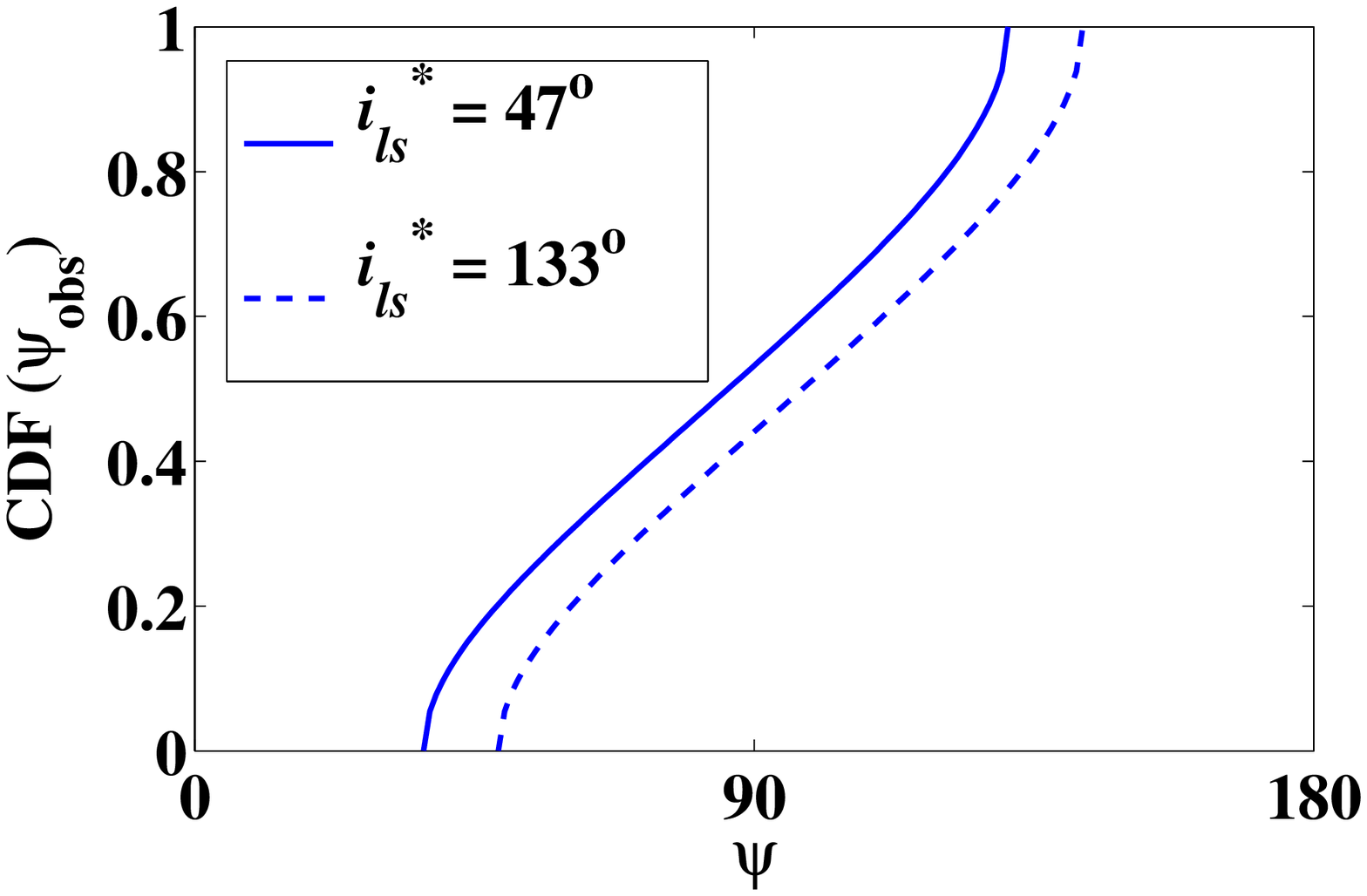}
\caption{{\it The cumulative distribution function of $\psi$}. This calculation is based on the observed parameters from  \citet{Huber+13}. We assume that the angle  between the stellar spin axis (${\bf n}_s$) and the normal to the innermost orbit (${\bf n}_{in}$) in the azimuthal direction around the line of sight (i.e., $\alpha$ in the schematic to the left) is random (taken from a uniform distribution). This enable us to produce a distribution function and not only a lower limit, see text for more details. We show a schematic of the geometry in the right panel. The solid curve corresponds to $i_{ls}^\star = 47^\circ\pm6$, and the dashed curve corresponds to $i_{ls}^\star = 133^\circ\pm6$ (due to the degeneracy in the asteroseismology measurements).}   \label{fig:psi} 
\end{center}
\end{figure*}

  \section{The Obliquity Distribution Function  }\label{sec:psi}

\citet{Huber+13}  analyzed the stellar oscillations observed in the {\it Kepler} photometry and used the splitting of the observed oscillation frequencies to measure the inclination between the stellar spin axis and the line of sight, finding $i_{ls}^\star = 47^\circ\pm6$. With the transit photometry, Huber et al. also measured the inclination of the inner planet's orbit with respect to the line of sight, finding $i_{ls}^{b} =83.84^\circ$$^{+0.26}_{-0.25}$. Together, these angles place a lower limit on the three-dimensional angle between the stellar spin axis and planetary orbital plane of $\psi>37^\circ$. 
%

 The angle between the normal of the orbit and the stellar spin is not simply $i_{ls}^{b}+i_{ls}^\star$ since, for example,  the angle, $i_{ls}^{b}$ can have different values on the sky plane (different values of $\alpha$ as shown in Figure \ref{fig:psi}). 
 In this simple geometrical configuration (see Figure~\ref{fig:psi}, left panel) and defining ${\bf L}_{in}$ and ${\bf S}$ as the angular momentum of the innermost orbit and stellar spin, respectively, the obliquity is defined by the scalar product between the three dimensional spin axis unit vector ${\bf{n}}_s = {\bf S}/S=(\sin i_{ls}^\star,0,\cos i_{ls}^\star)$ and the three dimensional normal to the innermost orbit ${\bf n}_{in}={\bf L}_{in} / L=(\sin i_{ls}^{b},0,\cos i_{ls}^{b})$, in random orientation with each other:
 
    \begin{equation}\label{Eq:psi}
 \cos \psi={\bf{n}}_s\cdot {\bf R}_{ls} {\bf{n}}_{in} \ .
 \end{equation}  
Here
    \begin{equation}\label{eq:Rz}
\bf{R}_{ls}  (\alpha) = \left(
\begin{array}{ccc}
\cos\alpha & -\sin\alpha & 0 \\
\sin\alpha & \cos\alpha & 0 \\
0 & 0 & 1
\end{array}\right)
\
 \end{equation}  
  is the rotation matrix in the azimuthal direction around the line of sight. 
We assume that $\alpha$, the angle  between the stellar spin and the orbital angular momentum in the azimuthal direction around the line of sight, is uniformly distributed. It is sufficient to  multiply only once  by the rotation matrix, with the random angle. 
Therefore, from Equation (\ref{Eq:psi}) we can estimate the cumulative distribution function of  $\psi$. As shown in the right panel of Figure \ref{fig:psi}, the lower limit on $\psi$ is of course the same one found by  \citet{Huber+13}, i.e,. $\psi>37^\circ$, but an upper limit of $131^\circ$ also exists and both these values have the same probability, which is larger than the probability of the angles in the range of $37^\circ<\psi<131^\circ$. We use $\psi_{obs}$ to denote the observationally constrained value of $\psi$. Note that due to the degeneracy in the asteroseismology measurements, $i_{ls}^\star$ could also be $133^\circ\pm6$. Setting $i_{ls}^\star = 133^\circ$, $\psi_{obs}$ is in the range of $49^\circ<\psi<143^\circ$ (see the dashed line in Figure \ref{fig:psi}). Therefore, adding these two pieces together, the distribution of $\psi_{obs}$ is symmetric over $37^\circ<\psi<143^\circ$. This distorts $\psi_{obs}$ only slightly, because $i_{ls}^{b}$ is almost $90^\circ$. Accordingly, we adopt $i_{ls}^\star = 47^\circ\pm6$, and have $\psi_{obs}$ constrained in the range of $37^\circ<\psi<131^\circ$ for the following discussion.
  
\section{Obliquity and inclination Evolution in the Presence of an Outer Perturber }

\subsection{Overview of the System Architecture }\label{sec:psievol}

In a sufficiently packed multi--planet system the planets' apsidal precessions are dictated by both the outer orbital companion and gravitational interactions between the two inner planets. In our case, the inner two planets are packed very close together, which suppresses eccentricity excitations that may arise due to the gravitational perturbations induced by the perturber (planet ``d"). If this perturber is inclined with respect to the orbital plane of the inner planets, then the plane will precess \citep[e.g.,][]{Inn+97,Takeda, Mardling10, Kaib11, Boue14}.  However the exact evolution of the obliquity and its current value are highly sensitive to the initial configuration of the system and, specifically, to the inclination of the outer orbit with respect to the inner one.

We first evolve the system with direct N-body integration using {\tt Mercury} software package \citep{Mercury} and then use our numerical results to evaluate the spin-orbit evolution (\S~\ref{sec:psiprec}). The latter is being set by the point mass dynamics (see below for more details).   
The system orbital parameters are set initially to 
 $a_b=0.1028$~AU, $a_c=0.1652$~AU \citep[based on the orbital solution provided by][]{Huber+13}. Since the properties of the outer body are yet unknown, we set $a_d = 2$~AU as an illustrative example following the dynamical simulation of Kepler 56 in \citet{Huber+13}. We work in the invariable plane where the z axis is parallel to the total angular momentum, ${\bf L_{tot}}$. Therefore, the inclinations of the orbits are defined with respect to the total angular momentum. In this frame, we set for simplicity $\omega_b=\omega_c=\omega_d=\Omega_b=\Omega_c=\Omega_d=0$, where $\omega_j$ ($\Omega_j$) is the argument of perihelion (longitude of ascending nodes) of the planet $j$. In addition, we simplify the system by imposing zero mutual inclination between the two inner planets and by setting the eccentricity of the two inner planets to zero \cite[which is consistent with ][estimate]{Huber+13}. 
 Following \citet{Huber+13}, we also take the mean anomalies to be  $f_b=57^\circ$, $f_c=182^\circ$
and $f_d=256^\circ$. 
The outer orbit eccentricity ($e_d$) does not affect the evolution of the system significantly, thus we only show results for $e_{d} = 0$.
The parameter that sets the system evolution is the mutual inclination between the outer planet's orbit and the inner plane, $i_{\rm mut}$, which we discuss in details below. Given the observed obliquity distribution (Figure  \ref{fig:psi}) we calculate next the probability distribution of the inclination of the system as a function of the system initial conditions.


 \subsection{Dynamics of Kepler 56}\label{sec:psiprec}
In the presence of a tilted outer orbit with inclination $i_{\rm mut}$, the two inner planets will precess around the total angular momentum vector. Note that the precession of the orbit due to  the oblateness of the star is negligible in this case. The torque felt by  planet ``b" due to stellar oblateness\footnote{The $J_2$ coefficient, which approximates the non-spherical shape by the star level of oblateness,  was calculated following \citet{1998EKH}.} is more than two orders of magnitude smaller than the torque due to planet ``c" (see \citealt{Tremaine+09} and  \citealt{Tamayo+13}). Therefore, the orbital evolution is not affected by the torque due to the stellar oblateness, and the system is in the ``pure orbital regime" \citep{Boue14}. We thus obtain the orbital evolution from an N-body simulation. 

The obliquity angle is defined with respect to the innermost planet's orbital angular momentum, ${\bf L}_b$. Thus, a natural coordinate choice for the spin is the Laplace--Runge--Lenz $({\bf \hat{q}}_b,{\bf \hat{h}}_b,{\bf \hat{e}}_b)$. Here, ${\bf \hat{e}}_b$ is the eccentricity vector (whose direction is toward the pericenter of planet ``b" orbit),  ${\bf \hat{h}}_b$ is the unit vector parallel to the orbital angular momentum of planet ``b" (the vector ${\bf h}_b$ is the specific angular momentum vector, i.e., ${\bf L}_b=m_\star m_b /(m_\star +m_b) {\bf h}_b$), and ${\bf \hat{q}}_b$ completes the right-hand triad of unit vectors. In this notation the precession of the stellar spin, ${\bf S}=(S_e,S_q,S_h)$, due to one planet is simply \citep{1998EKH} 
\begin{equation}\label{eq:dsdt}
\frac{d{\bf S}}{dt}_{prec,a}={\bf S}\times {\bf K}_b + \frac{m_\star m_b}{m_\star +m_b} h_{b}/I_2 (-\tilde{Y}_b {\bf\hat{e}}_b+\tilde{X}_b{\bf\hat{q}}_b+\tilde{W}_b{\bf\hat{h}}_b) \ ,
\end{equation}
where $h_{b}=[G(m_\star+m_b)a_b(1+e_b^2)]^{1/2}$, $G$ is the gravitational constant, and ${\bf K}_b=(X_b,Y_b,Z_b)$ represents the precession due to the orbital evolution:
\begin{eqnarray}
X_b&=& \frac{di_b}{dt} \cos \omega_b +\frac{d\Omega_b}{dt}\sin \omega_b \sin i_b \ , \\
Y_b&=& -\frac{di_b}{dt}\sin \omega_b + \frac{d\Omega_b}{dt}\cos \omega_b \sin i_b \ , \\
Z_b&=& \frac{d \omega_b}{dt} + \frac{d\Omega_b}{dt} \cos i_b \ , 
\end{eqnarray}
and $\tilde{X}_b$, $\tilde{Y}_b$ and $\tilde{W}_b$ represent the torque due to the stellar oblateness and the tidal dissipation:
\begin{eqnarray}
\tilde{X}_b&=&  -\frac{m_b k_\star R_\star^5}{\mu \dot{l} a_b^5} \frac{S_h S_e}{(1-e_b^2)^2} \\ \nonumber
&-&\frac{S_q}{2 \dot{l} t_{F\star}} \frac{1+(9/2)e_b^2 + (5/8)e_b^4}{(1-e_b^2)^5}\ , \\
\tilde{Y}_b&=&  -\frac{m_b k_\star R_\star^5}{\mu \dot{l} a_b^5} \frac{S_h S_q}{(1-e_b^2)^2} \\ \nonumber
&-& \frac{S_e}{2 \dot{l}t_{F\star}}\frac{1+(9/2)e_b^2 + (5/8)e_b^4}{(1-e_b^2)^5}\ \label{eq:W} , \\ 
\tilde{W}_b &=& \frac{1}{t_{F\star}}\Big[\frac{1+(15/2)e_b^2+(45/8)e_b^4+(5/16)e_b^6}{(1-e_b^2)^{13/2}} \\ \nonumber
&-& \frac{S_h}{\dot{l}}\frac{1+3e_b^2+(3/8)e_b^4}{(1-e^2)^5}\Big] \ , \label{eq:prec}
\end{eqnarray}
where $\dot{l} = \sqrt{G m_\star / a_b^3}$, and
 \begin{equation}
t_{F\star} = \frac{t_{V\star}}{9} \frac{m_\star^2}{(m_\star+m_b)m_b} \left(\frac{a_b}{R_\star}\right)^8 \frac{1}{(1+2k_2)^2} \ , \label{eq:TF}
 \end{equation}
 
To calculate the orbital evolution due to the orbital precession (the ${\bf K}_b$ term), we take the time evolution of $\omega,\Omega$ and $i$ of planets ``b"  directly from the N-body integration. This dominates the obliquity variation. The tidal effects are negligible until planet b is almost engulfed (see discussion on the future evolution of Kepler-56 in \textsection \ref{sec:Sevol}). The timescale for the evolution of planet b's orbital separation due tidal dissipation in the star is defined in terms of the stellar viscous timescale $t_{V\star}$. $t_{V\star}$ is set to be 50~yr and kept constant, where $t_{V\star}$ corresponds to $Q\sim10^6$ for a 10 day orbit.
The parameter $k_2$ is the apsidal precession constant, which is related to the Love parameter $k_{L}$ via $k_2=2k_L$ ({a similar equation exists for planet ``b" and ``c"}).  
 Note that the effects of tides in the planets are negligible. In fact, assuming a viscous timescale corresponding to $Q=12$ and $10^5$ for planet  ``b" and ``c" \citep{MD00}, respectively, 
 the small planets radii yield much longer tidal timescales [see equation (\ref{eq:TF})].  In any case, the unconstrained nature of exoplanets makes it difficult to conclude how their tidal coefficients evolve. 
 

\begin{figure}[!t]
\hspace{-0.7cm}
\includegraphics[width=10cm,clip=true]{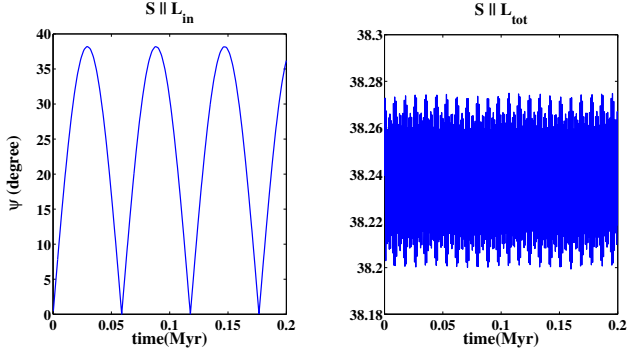}
\caption{{\bf Short time scale obliquity  evolution for the two scenarios.} The left panel shows the evolution in the ${\bf S} || {\bf L_{in}}$ with an initial  $i_{\rm mut}=20^\circ$ scenario while the right panel is for the ${\bf S} || {\bf L_{tot}}$ with an initial $i_{\rm mut}=40^\circ$. The orbital evolution was done using direct N-body integration. }
   \label{fig:obliq} 
\end{figure}

{Equations (\ref{eq:dsdt})--(\ref{eq:prec}) imply that the time evolution of $i_{\rm mut}$ (and thus $\psi$) depends on the initial system's configuration. This can be constrained from the observed obliquity distribution}. In Figure~\ref{fig:obliq} we show the evolution of $\psi$ assuming two possible initial configurations: ${\bf S}$ parallel to ${\bf L_{in}}$ and $i_{\rm mut}=20^{\circ}$ (left, hereafter ``${\bf S} || {\bf L_{in}}$" scenario), and ${\bf S}$ parallel to ${\bf L_{tot}}$ and $i_{\rm mut}=40^{\circ}$ (right, hereafter ``${\bf S} || {\bf L_{tot}}$" scenario), where ${\bf L_{in}}$ and $\bf L_{tot}$ are the orbital angular momentum of the {\it inner} two planets and the {\it total} orbital angular momentum, respectively. We show below that these values for $i_{\rm mut}$ give a misalignment of at least $37^\circ$ during the evolution (the minimum value constrained observationally). 

In the ${\bf S} || {\bf L_{in}}$ scenario, $\psi$ oscillates between well-aligned ($\psi=0^\circ$) and $\sim2\times i_{\rm mut}$ ($\sim 38.2^\circ$). In this case, we postulate that the system formed initially  in a disk and planet ``d"  was perhaps scattered to large inclinations \citep[e.g.,][]{RF96}, causing the obliquity angle to precess between $0^\circ$ and $\sim2\times i_{\rm mut}$. Another possible case for this configuration is accretion of material onto the protoplanetry disk, which can tilt the outer parts of the disk and the total angular momentum \citep{Bate10, Tremaine11, Thies+11}. Therefore, in the ${\bf S} || {\bf L_{in}}$ scenario, $\psi\sim 40^\circ$ can be produced by an initial inclination $i_{\rm mut} > 20$. Note that a retrograde configuration with $i_{\rm mut}=160^\circ$ can also produce  $\psi\sim 40^\circ$.

In the ${\bf S} || {\bf L_{tot}}$ scenario, $\psi$ remains close to the initial value. This configuration could have occurred if the inner parts of the disk were warped perhaps due to magnetic interactions with the inner disk edge \citep[e.g.,][]{Lai+10}.
Therefore in the ${\bf S} || {\bf L_{tot}}$ scenario, $\psi\sim 40^\circ$ can be produced by an initial $i_{\rm mut}=40^\circ$. 

We show below that for the ${\bf S} || {\bf L_{in}}$ scenario $\psi$ is more likely to be detected in the maximum (at $\sim 38.2^\circ$) where the derivative is closer to zero. For each possible  obliquity value  $\tilde{\psi}\in(0^\circ,180^\circ)$, we  derived a cumulative distribution function of the mutual inclination, where CDF$(\tilde{\psi}|i_{\rm mut})=\Delta t(\psi<\tilde{\psi}|i_{\rm mut})/t$, where $\Delta t$ is the time interval. This quantity will be used below to estimate the probability distribution of the system configuration for the actual observations.  We run 35 N-body runs, for an array of initial inclinations $i_{\rm mut}$ between $5^\circ$ and $175^\circ$, and calculate the cumulative probability for the two scenarios.

\subsection{Inferring the Inclination Distribution Function from Observations}\label{sec:pofi}

\begin{figure}[t]
\includegraphics[width=10cm,clip=true]{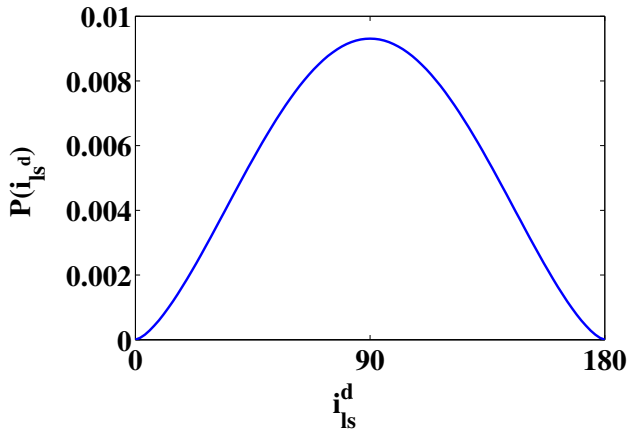}
\caption{{\bf The probability distribution of $i_{ls}^{d}$}. We calculate this probability assuming $m_d$ follows the mass function of \citet{Cumming+08}. \label{fig:proid} } 
\end{figure}

\begin{figure*}[t]
\begin{center}
\includegraphics[width=\linewidth]{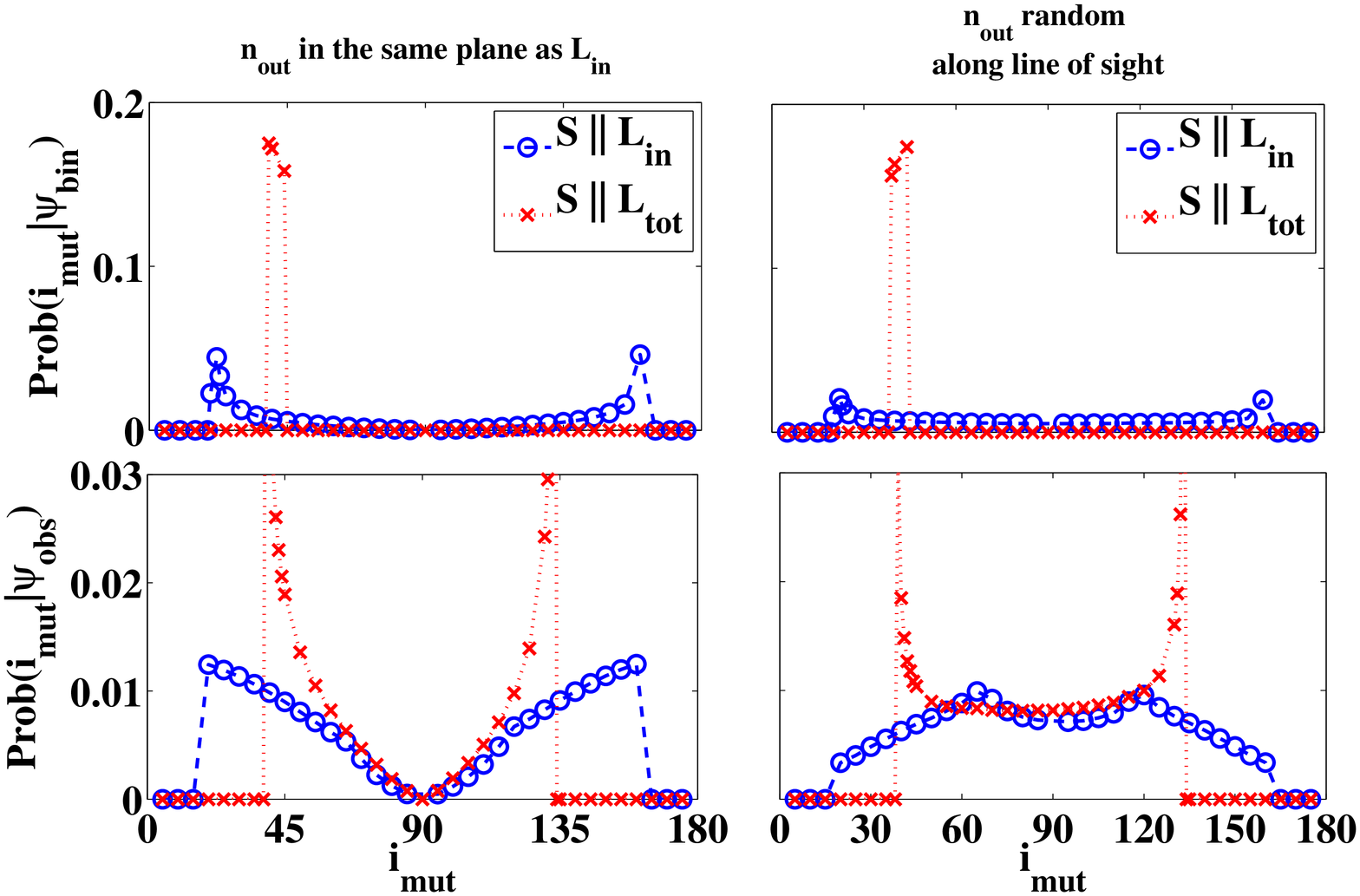}
\caption{{\bf The probability distribution of the mutual inclination inferred from observations.} We consider the two scenarios  ${\bf S} || {\bf L_{in}}$  (blue circle) ${\bf S} || {\bf L_{tot}}$ (red lines), and two possible probability distribution on $p(i_{\rm mut})$. The {\it left panels} are {\bf for ${\bf n}_d$ lying in the plane defined by ${\bf n}_{in}$ and the line of sight}, i.e., $i_{\rm mut}=i_{ls}^{b}-i_{ls}^d$, while in the {\it right panels} we assume random orientation (see text). The {\it top panels} show a specific example for the advantage in having a more precise observation $\psi_{bin}=37-43^\circ$, and the {\it bottom panels} show the results form the observed cumulative distribution (Figure \ref{fig:psi}).  \label{fig:inclprob} }\end{center} 
\end{figure*}

When the spin-orbit misalignment is due to the dynamical interaction between the planets, the obliquity distribution function derived from observations (Section \ref{sec:psi}, Figure \ref{fig:psi}) can be used to place strong constrains on the mutual inclination between the inner planets and planet ``d", i.e., $i_{\rm mut}$. 
We calculate the conditional probability distribution of $i_{\rm mut}$ given the observed distribution $\psi_{obs}$, i.e., $p(i_{\rm mut}|\psi_{obs})$. This  posterior probability can be written as
\begin{equation}\label{eq:Bayes}
p(i_{\rm mut}|\psi_{obs})=\frac{p(\psi_{obs}|i_{\rm mut}) p(i_{\rm mut})}{p(\psi_{obs})} \ ,
\end{equation}
where $p(\psi_{obs})$  is a normalization term, which we disregard because the shape of the distribution is of larger significance than the absolute probability here, and the absolute probability is out of the scope of this paper. 

Furthermore, we  use the distribution function of planet ``d" line of sight inclination, $i_{ls}^d$, to estimate the  prior probability, $p(i_{\rm mut})$. Note, that the actual value of  $m_d \sin i_{ls}^d $ only affect the normalization of the probability, but since we care about the shape of the probability we can ignore this. Note that if we assume the outer orbit to be isotropically distributed, the probability density function for $i_{ls}^d$ takes the form of $\sin{i}_{ls}^d$. This suggests that the most probable value for $i_{ls}^d$ is $90^\circ$.

Following \citet{Ho+11} we calculate the probability  $p(i_{ls}^d)$ assuming  \citet{Cumming+08}  mass function for $m_d$ (see Figure \ref{fig:proid}). Note that the distribution in \citet{Cumming+08} is for $m\sin{i}$, not $m$. However, since the power law index is large, we use this power law for the mass distribution according to \citet{Ho+11}.
The angle we are actually interested in is  the angle between the normal to the outer orbit ${\bf n}_{out}$ and the normal to the inner orbit ${\bf n}_{in}$. While $i_{ls}^{b}$ has been measured to be $83.84^\circ$,  \citep{Huber+13} we have no information about the orientation of these two vectors  on the plane of the sky. Consider first the case where the three dimensional normal to the outer orbit, ${\bf n}_d$ lies in the plane defined by ${\bf n}_{in}$ and the line of sight. This yields a simple relation between the different angles, i.e., $i_{\rm mut}=i_{ls}^{b}-i_{ls}^d$. Therefore, $p(i_{\rm mut})=p(i_{ls}^{b}-i_{ls}^d)$, where the latter is calculated from $p(\sin i_{ls}^d)$ following \citet{Ho+11}.  Their mass distribution function  yields small $m_d$   (compared to the measured  $m_d \sin i_{ls}^d$), thus $\sin i_{ls}^d$ is more likely to be close to its maximum of 1. This suggests angles near 90 degrees for $i_{ls}^d$, which implies that  $i_{\rm mut}=i_{ls}^{b}-i_{ls}^d$ is more likely to have a small value. 

However, another possible prior is that  ${\bf n}_d$ has a random orientation (similar to the configuration depicted in the left side of Figure \ref{fig:psi}). Thus, as  in Section \ref{sec:psi}, we multiply the normal to the orbit with the rotation matrix in Eq.~(\ref{eq:Rz}) assuming a random azimuthal angle $\alpha$,  i.e., 
\begin{equation}\label{eq:cosi}
\cos i_{\rm mut} = {\bf n}_{out}( i_{ls}^d) \cdot {\bf R}_{ls}(\alpha) {\bf n}_{in}(i_{ls}^{b})  \ ,
\end{equation}
 where ${\bf n}_{out}( i_{ls}^d)$ is chosen with $p(\sin i_{ls}^d)$ distribution, which gives $p(i_{\rm mut})$. This prior also gives a high probability for large values of $i_{\rm mut}$, as this case covers large parts of the parameter space.
 Below we consider these two cases.


The probability of $\psi_{obs}$ for a given $i_{\rm mut}$, i.e., $p(\psi_{obs}|i_{\rm mut})$ can be calculated from
\begin{equation}\label{eq:pobsofi}
p(\psi_{obs}|i_{\rm mut}) = \int_0^\infty p(\tilde{\psi}|i_{\rm mut}) p_{obs}(\tilde{\psi}) d\tilde{\psi} \ ,
\end{equation}
where $ p_{obs}(\psi)$ was computed in Section \ref{sec:psi}, Figure \ref{fig:psi}. The probability $p(\psi|i_{\rm mut})$ is calculated from theory for the two different cases, i.e., ${\bf S} || {\bf L_{tot}}$  and  ${\bf S} || {\bf L_{in}}$. In the discrete description we calculate the probability distribution of $i_{\rm mut}$ for each $\psi$. This can be easily derived from the cumulative distribution function calculated in Section \ref{sec:psi}, Figure \ref{fig:psi}.

Using Equations (\ref{eq:Bayes}) $-$ (\ref{eq:pobsofi}) we can find the mutual inclination probability function given the observed obliquity distribution. This is depicted in Figure \ref{fig:inclprob}, {\it bottom panels}.  We consider the two initial configurations scenarios, i.e., ${\bf S} || {\bf L_{tot}}$  and  ${\bf S} || {\bf L_{in}}$, and  the two $p(i_{ls}^d)$ cases, i.e.,  ${\bf n}_{out}$ random along line of sight ({\it right panels}) and ${\bf n}_{out}$ in the same plane as ${\bf n}_{in}$ ({\it left panels}).  Since the  obliquity distribution function derived from observations has  two high probability peaks, ($\psi=37^\circ$ and $\psi=131^\circ$), the possible $i_{\rm mut}$ values that can produce such distribution function also have two peaks. In the case of ${\bf S} || {\bf L_{in}}$, the double peak distribution is also probable since the precession of a retrograde orbit can as well produce this configuration. Note that if we also consider the case when $i_{ls}^\star = 133^\circ\pm6$ (due to the degeneracy in the asteroseismology measurements), $\psi_{obs}$ is symmetric, and $p(i_{\rm mut}|\psi_{obs})$ would also be symmetric.

Interestingly,  better observations may help constraining $i_{\rm mut}$ but will not disentangle the degeneracy between the ${\bf S} || {\bf L_{tot}}$  and  ${\bf S} || {\bf L_{in}}$ cases. We show this in the top panels of Figure \ref{fig:inclprob}, where we consider an example of $\psi=40\pm3^\circ$. In the ${\bf S} || {\bf L_{tot}}$ scenario, the symmetry is broken, since, there is a direct link between the obliquity and $i_{\rm mut}$ in this case, as seen from the right panel of Figure 2. Note that the two different $p(i_{\rm mut})$ cases produce slight differences in the probability peak.   Assuming that  ${\bf n}_{out}$ and ${\bf n}_{in}$ are coplanar produces a decreasing probability toward  $i_{\rm mut}\sim 45^\circ$, as in this case near polar configurations are less likely. On the other hand, assuming a random orientation for ${\bf n}_{out}$ produces an increasing probability toward the larger $i_{\rm mut}$ values. In fact, as mentioned above, this case yields a larger parameter space for near polar configurations. Having a precise observation also improves the $i_{\rm mut}$ estimation for the ${\bf S} || {\bf L_{in}}$  but the double peak probability remains, because the same obliquity can be reached in a prograde and retrograde configurations. 
The degeneracy can be broken only for the case where ${\bf S} || {\bf L_{tot}}$, a more precise measurement of $\psi$ will be available. This can be seen in the top panels of Figure \ref{fig:inclprob}, where $\psi_{bin}$ represents $37^\circ < \psi < 43^\circ$.

So far, we have assumed two possible priors for $p(i_{\rm mut})$. These represent two extreme possibilities, one which favors  low mutual inclinations and one which favors large values. The truth may lay in between. Thus, we have tested the possibility that  ${\bf n}_d$ is randomly oriented within a small interval as a prior (see the left side of Figure  \ref{fig:psi}, where $\alpha$, is now confined to a certain interval). In this case, differently from what figure 3 shows, we assumed an initial tilt of  $37^\circ$  between the stellar spin axis and the angular momentum of the inner orbit. This way, we consider the possibility that the source of the obliquity is not dynamical. We find that for $\alpha\gsim10^\circ$ equation (\ref{eq:Bayes}) and the observed obliquity distribution favors large mutual inclinations. In other words, the three planets will be aligned, and the observations will be consistent with tilt of the star or the disk in the migration scenario, if the random angle $\alpha<10^\circ$


 \section{Tidal and Stellar Evolution}\label{sec:Sevol}
\begin{figure*}
 \begin{center}
\includegraphics[width=5.5in, height=3in]{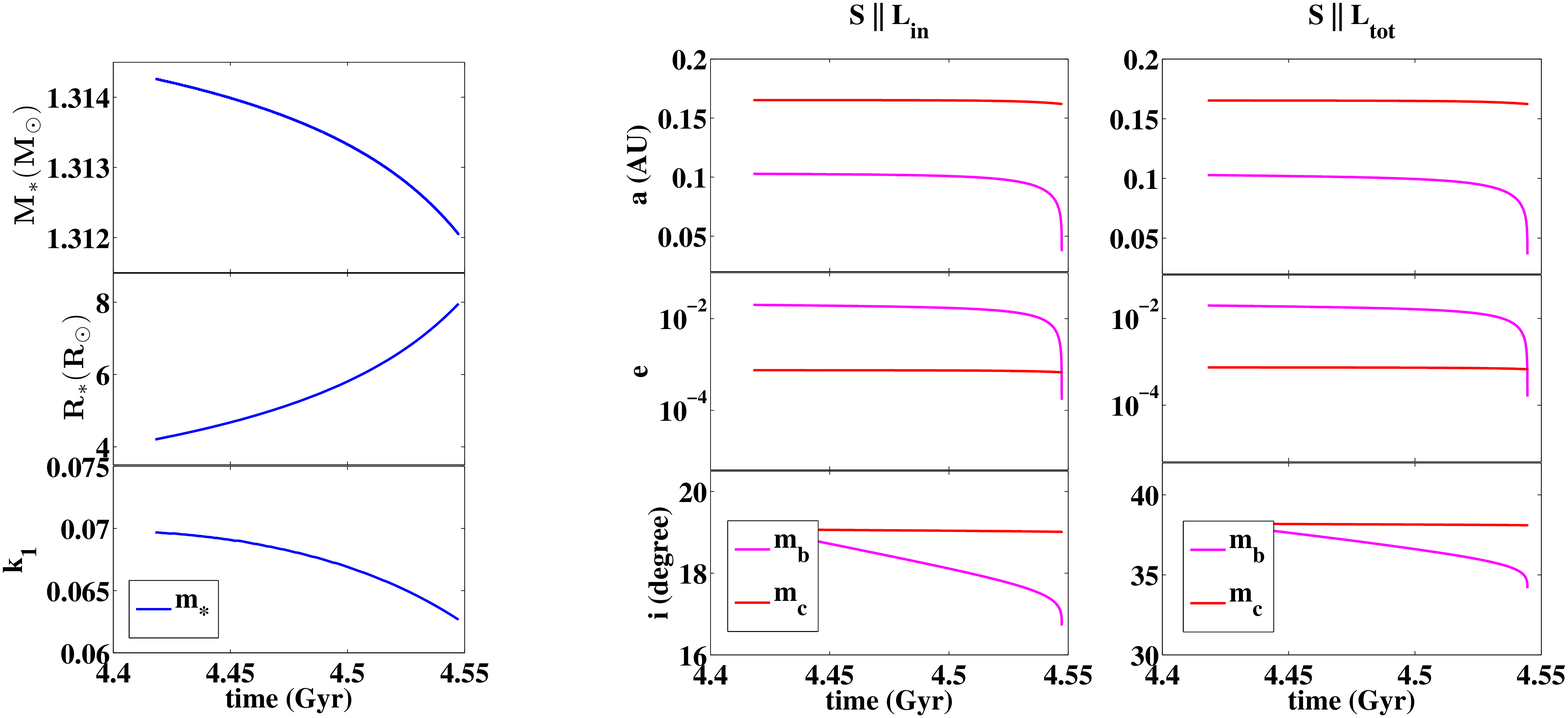} 
\caption{{\bf Future evolution of the star and the innermost planets.} {\it Left column (from top to bottom):} evolution of the stellar mass, radius and the apsidal precession constant ($k_2$) computed with MESA \citep{Paxton+11,Paxton+13}. {\it Middle and right columns (from top to bottom)}: evolution of the semi-major   axis, eccentricity, and inclination with respect to the invariable plane for planet ``b" (magenta lines) and planet ``c" (red lines). In the middle panels we consider the  ${\bf S} || {\bf L_{in}}$ scenario with an initial  $i_{\rm mut}=20^\circ$, while in the right panels we consider the  ${\bf S} || {\bf L_{tot}}$ scenario with an initial  $i_{\rm mut}=40^\circ$. We start the calculation at the present time and we stop it when the innermost planet is engulfed ($a_b = R_\star$). The evolution depicted is due to tidal interactions between the evolving star and the two inner planets, also accounting for the point mass dynamics via direct N-body integrations. 
}
\label{fig:evolsum}
\end{center}
\vspace{0.1cm}
\end{figure*}

Here we focus on the fate of the innermost planet and the future evolution of the obliquity as a result of tidal dissipation in the star and stellar evolution.
  We compute a detailed model of the host star with the publicly available stellar evolution code {\tt MESA} \citep[version  4798][]{Paxton+11,Paxton+13}. Specifically, we follow \citet{Huber+13} and consider a star with an initial mass and metallicity of $1.32$~M$_{\odot}$ and $Z = 0.032$, respectively. We evolve the stellar model with the same physical assumptions adopted in \cite{ValsecchiRasio14a}. Briefly, we account for stellar wind mass loss following the test suite example provided with {\tt MESA}  for the evolution of a $1$~M$_{\odot}$ star, and we set the mixing length $\alpha_{MLT}$ parameter to $1.92$, following the {\tt MESA}  star Standard Solar Model \citep[][Table 10]{Paxton+11}.
Note that the mass loss is negligible in this case, since the star is only slightly evolved (as shown in Figure \ref{fig:evolsum}). This negligible mass loss explains why the planets' orbits are significantly expanding, differently from the case of the Earth when the Sun evolves into a red giant. The model agrees with the observationally inferred stellar mass, radius, and effective temperature (within 1$\,\sigma$) at $4.418$~Gyr. The latter is consistent with the age quoted by \citet{Huber+13} within $1\sigma$ ($3.5\pm 1.3$~Gyr).

{The advanced evolutionary stage of the host star (which is off its Main Sequence) poses the interesting possibility that, if Kepler-56 is similar to other Kepler multi-planet systems, it may have had planets that were engulfed as the star expanded \citep[such possibilities have been investigated in the literature,][]{Bear+11,Bear+12}. If this is the case, the observed small stellar rotation rate suggests that the host star in Kepler-56 did not engulf a large planet. In fact, to increase the stellar spin by more than $10\%$, the engulfed planet should have had a mass larger than $0.6$~M$_J$ (neglecting the possibility of core-envelope decoupling, see, e.g., \citealt{Teitler+14}).  However, we note that  magnetic braking, stellar winds, and the expansion of the star as a result of natural stellar evolution might all contribute to spin down after engulfment}. {Nevertheless, it also seems unlikely (but not impossible) that a very massive planet could have migrated to the innermost configuration, with two lighter planets outside (planets ``b" and ``c") which also supports the notion that no inner planet was engulfed.}
 

%
%
The inner planets' orbital evolution is affected by tides, whose efficiency changes as the star evolves. In the left  panels of Fig.~\ref{fig:evolsum} we show the forward evolution of the stellar mass, radius, and Love parameter. The latter was computed following \cite{Valsecchi+12}.

Both the specific angular momentum ($h_{b}$) and the eccentricity undergo tidal dissipation, which leads to circularization and orbital shrinking. 
Following \citet{1998EKH} we have
\begin{eqnarray}
\frac{de_b}{dt}&=&-\tilde{V}_b e_b \ , \\
\frac{dh_b}{dt} &=& -\tilde{W}_b h_b \ .
\end{eqnarray}
The parameters $\tilde{W}_b$ and $\tilde{V}_b$ are the dissipation coefficients \citep[see][]{1998EKH}, where $\tilde{W}_b$ is given by equation (\ref{eq:W}), and $\tilde{V}_b$ is defined as:
\begin{eqnarray}
\tilde{V}_b &=& \frac{9}{t_{F\star}}\Big[\frac{1+(15/4)e_b^2+(15/8)e_b^4+(5/64)e_b^6}{(1-e_b^2)^{13/2}} \\ \nonumber
&-& \frac{11S_h}{18\dot{l}}\frac{1+(3/2)e_b^2+(1/8)e_b^4}{(1-e^2)^5}\Big]\ . \label{eq:V} 
\end{eqnarray}

We compute the evolution of the orbital separation, eccentricity, and inclination, using the extrapolated orbital parameters from the initial direct N-body integration, together with the equations mentioned above. We stop the integration when the innermost planet is engulfed ($a_b = R_\star$) and neglect possible mass transfer events between the planet and the star (e.g., \citealt{TrillingBGLHB1998}), for simplicity.
The evolution is shown in the right two panels of Figure \ref{fig:evolsum}. 
During the first $\sim 0.1$~Gyr of evolution, the star loses about $0.1\%$ of its mass and its radius expands by about $40\%$. After this stage tidal effects become increasingly important and planet ``b" is quickly engulfed. We note that the tidal treatment adopted here does not fully account for how the evolution of the star affects the efficiency of tides (i.e. the stellar viscous timescale $t_{V\star}$ is kept fixed). However, a more consistent orbital evolution calculation with the method adopted in \cite{ValsecchiRasio14b,ValsecchiRasio14a}, but only accounting for the evolution of the innermost planet, yields similar results. Note also that the precession due to the stellar oblateness affects the final stages of the evolution (very close to the final engulfment of planet ``b"). This occurs because tidal dissipation dominates the dynamics only towards the end of the evolution right before the engulfment, and thus, it does not change the overall orbital dynamics. Moreover, we find that the final semimajor axis of the planet is $\sim 0.03$ AU during the engulfment, which is within twice the Roche limit (the Roche limit is $0.016$ AU according to the prescription of \citet{Paczyski71}). This suggests that the planet may be tidally distorted during the engulfment and that the accumulated heat due to tidal dissipation as the planet orbits the star multiple times may increase the chance of tidal disruption \citep{Li13}. Past studies have investigated the engulfment of planets by their host stars \citep{Nordhaus10, Bear11, Kaib11, Kratter12, Veras13, Lillo-Box14}. Figure \ref{fig:evolsum} shows that the innermost planet will be engulfed in $\sim129$ Myr. Similarly, the second planet (Kepler 56c) will be engulfed in less than $\sim155$ Myr.

 %
%
 
\begin{figure}
\hspace{-0.5cm}
\includegraphics[width=9.7cm,clip=true]{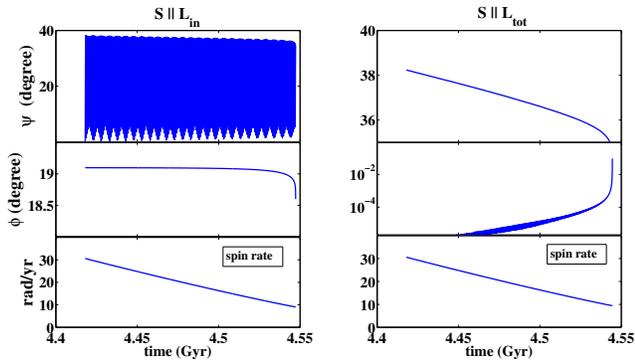} 
\caption{ {\bf The evolution of the stellar spin}. {\it Top: } obliquity; {\it middle: } angle between the stellar spin and the total angular momentum; {\it bottom:} spin magnitude $|{\bf S}|$ in units of rad~yr$^{-1}$. The initial spin period is $75$~d, which translates to a spin rate of $\sim 30$~rad~yr$^{-1}$. The left panel shows the evolution in the ${\bf S} || {\bf L_{in}}$ scenario with an initial $i_{\rm mut}=20^\circ$, while the right panel is for the ${\bf S} || {\bf L_{tot}}$ scenario with an initial $i_{\rm mut}=40^\circ$. }\label{fig:spinsum}
\vspace{0.1cm}
\end{figure}
 
The tidal evolution of the inner planets affects the stellar spin evolution (equation (\ref{eq:dsdt})). The same equation holds for planet ``c" (substituting subscript ``b" with ``c"). The stellar spin evolves due to the precession of planets ``b" and ``c", and their tidal torque. 
{ We extrapolate the evolution of their precession directly from the N-body calculation.}
The evolution of the stellar spin direction and magnitude is shown in Figure~\ref{fig:spinsum}. In particular, we show the evolution of the obliquity $\psi$ and the angle between the stellar spin and the total angular momentum ($\phi$).  The magnitude of the spin decreases due to the mass lost and the expansion of the stellar radius (irrespective of the scenario considered).  This exercise reveals that the obliquity behavior for the two cases does not vary much as the star evolves. In the 
 ${\bf S} || {\bf L_{in}}$ scenario with an initial  $i_{\rm mut}=20^\circ$, the obliquity oscillates between zero and $\sim 40^\circ$, the amplitude slightly decreases, and additional modulations due to tides appears. In the  ${\bf S} || {\bf L_{tot}}$ scenario with an initial $i_{\rm mut}=40^\circ$, the obliquity monotonically decreases.  
 
 \section{Discussion}\label{sec:dis}
 
 We studied the configuration and obliquity of Kepler-56, a multi planet system with two coplanar  inner planets that are misaligned with respect to their host star. 
Two main scenarios were proposed in the literature to explain the large obliquities observed for close-in exoplanets. The first model involves dynamical evolution between multi planetary members or stellar companion  \citep[e.g.,][]{Winn+10b, Dan,Sourav+08,Nag+08,Naoz11,Naoz+12bin, Naoz13, WY11,Li+13}. The second model proposes disk migration as the main mechanism which controls the planetary configuration, while the star spin axis is tilted with respect to the planets by other mechanisms \citep[e.g.][]{Winn+10b,Lai+10,Rogers+12,Rogers+13IGW, Spalding14}. The two scenarios lead to different configurations for the configuration of the planets with respect to each other and the star. The dynamical scenario predicts that large obliquities are associated with an inclined perturber, while the disk-migration scenario predicts aligned planetary systems. 
We showed that the large obliquity observed in Kepler-56 is consistent with a dynamical nature.

 
 We showed that we can improve the \citet{Huber+13} lower limit on the obliquity ($\psi>37^\circ$). Specifically, using the  \citet{Huber+13} current  observations, we  found the probability distribution of the observed true obliquity (see Figure \ref{fig:psi}). This probability has a  large range with two main peaks at $\psi=37^\circ$ and $\psi=131^\circ$. Furthermore, using this probability distribution we gave the probability distribution of the inclination of the third planet with respect to the inner two ($i_{\rm mut}$). This is highly dependent on the system's initial conditions. For this reason, we explored two scenarios: ${\bf S} || {\bf L_{in}}$ and ${\bf S} || {\bf L_{tot}}$. In the former, the initial spin axis of the star was set along the orbital angular momentum of the inner two planets. A possible origin for this configuration is  that  the system formed initially  in a disk and  the third Jupiter-like planet  was perhaps scattered to a large inclination. Instead, in the ${\bf S} || {\bf L_{tot}}$ scenario, the initial stellar spin was set parallel to the {\it total} orbital angular momentum. This initial condition may be ad hoc, and possibly caused by, e.g., magnetic interactions \citet[e.g.,][]{Lai+10} warping the inner parts of the disk. For these two scenarios, we found the  mutual inclination probability function for the observed obliquity distribution (see Figure \ref{fig:inclprob} {\it bottom panels}). Both configurations have a double peak distributions, with  zero probability of having aligned configuration between the two orbits. The degeneracy between the two probability peaks may be broken only for the ${\bf S} || {\bf L_{tot}}$ case, with a more precise measurement of $\psi$. However, a precise  measurement of $\psi$ would not  disentangle between the ${\bf S} || {\bf L_{tot}}$  and  ${\bf S} || {\bf L_{in}}$ cases, as shown in the top panels of Figure \ref{fig:inclprob}. 
 
 We finally considered the effect of the stellar evolution on the system's parameters and, specifically, the obliquity. We evolved the host star using  {\tt MESA} \citep{Paxton+11,Paxton+13} and extrapolated the planets orbital evolution calculated with direct N-body integration (since the latter is rather regular and periodic).  We have also included the spin precession and tidal evolution.  This exercise revealed that the obliquity behavior for the two cases does not vary significantly as the star evolves. It also shows that planet ``b" will be engulfed in $\sim 129$ Myr. 
 
 \section*{Acknowledgments}
SN is  supported by NASA through an Einstein Post--doctoral Fellowship awarded by the Chandra X-ray Center, which is operated by the Smithsonian Astrophysical Observatory for NASA under contract PF2-130096. This work was also supported by NASA Grant NNX12AI86G at Northwestern University. JAJ gratefully acknowledges funding from the Alfred P. Sloan and David \& Lucile Packard foundations.  
 
\bibliographystyle{hapj}

\bibliography{dynam}

\begin{thebibliography}{69}
\expandafter\ifx\csname natexlab\endcsname\relax\def\natexlab#1{#1}\fi

\bibitem[{{Albrecht} {et~al.}(2012){Albrecht}, {Winn}, {Johnson}, {Howard},
  {Marcy}, {Butler}, {Arriagada}, {Crane}, {Shectman}, {Thompson}, {Hirano},
  {Bakos}, \& {Hartman}}]{Albrecht+12}
{Albrecht}, S. {et~al.} 2012, \apj, 757, 18, 1206.6105

\bibitem[{{Albrecht} {et~al.}(2013){Albrecht}, {Winn}, {Marcy}, {Howard},
  {Isaacson}, \& {Johnson}}]{Albrecht+13}
{Albrecht}, S., {Winn}, J.~N., {Marcy}, G.~W., {Howard}, A.~W., {Isaacson}, H.,
  \& {Johnson}, J.~A. 2013, \apj, 771, 11, 1302.4443

\bibitem[{{Bate} {et~al.}(2010){Bate}, {Lodato}, \& {Pringle}}]{Bate10}
{Bate}, M.~R., {Lodato}, G., \& {Pringle}, J.~E. 2010, \mnras, 401, 1505,
  0909.4255

\bibitem[{{Batygin}(2012)}]{Batygin12}
{Batygin}, K. 2012, \nat, 491, 418

\bibitem[{{Bear} \& {Soker}(2011{\natexlab{a}})}]{Bear+11}
{Bear}, E., \& {Soker}, N. 2011{\natexlab{a}}, \mnras, 414, 1788, 1102.5487

\bibitem[{{Bear} \& {Soker}(2011{\natexlab{b}})}]{Bear11}
------. 2011{\natexlab{b}}, \mnras, 414, 1788, 1102.5487

\bibitem[{{Bear} \& {Soker}(2012)}]{Bear+12}
------. 2012, \apjl, 749, L14, 1202.1168

\bibitem[{{Bou{\'e}} \& {Fabrycky}(2014)}]{Boue14}
{Bou{\'e}}, G., \& {Fabrycky}, D. 2014, ArXiv e-prints, 1405.7636

\bibitem[{{Bowler} {et~al.}(2010){Bowler}, {Johnson}, {Marcy}, {Henry}, {Peek},
  {Fischer}, {Clubb}, {Liu}, {Reffert}, {Schwab}, \& {Lowe}}]{Bowler10}
{Bowler}, B.~P. {et~al.} 2010, \apj, 709, 396, 0912.0518

\bibitem[{{Chambers} \& {Migliorini}(1997)}]{Mercury}
{Chambers}, J.~E., \& {Migliorini}, F. 1997, in Bulletin of the American
  Astronomical Society, Vol.~29, AAS/Division for Planetary Sciences Meeting
  Abstracts \#29, 1024--+

\bibitem[{{Chaplin} {et~al.}(2013){Chaplin}, {Sanchis-Ojeda}, {Campante},
  {Handberg}, {Stello}, {Winn}, {Basu}, {Christensen-Dalsgaard}, {Davies},
  {Metcalfe}, {Buchhave}, {Fischer}, {Bedding}, {Cochran}, {Elsworth},
  {Gilliland}, {Hekker}, {Huber}, {Isaacson}, {Karoff}, {Kawaler}, {Kjeldsen},
  {Latham}, {Lund}, {Lundkvist}, {Marcy}, {Miglio}, {Barclay}, \&
  {Lissauer}}]{Chaplin+13}
{Chaplin}, W.~J. {et~al.} 2013, \apj, 766, 101, 1302.3728

\bibitem[{{Chatterjee} {et~al.}(2008){Chatterjee}, {Ford}, {Matsumura}, \&
  {Rasio}}]{Sourav+08}
{Chatterjee}, S., {Ford}, E.~B., {Matsumura}, S., \& {Rasio}, F.~A. 2008, \apj,
  686, 580, arXiv:astro-ph/0703166

\bibitem[{{Cumming} {et~al.}(2008){Cumming}, {Butler}, {Marcy}, {Vogt},
  {Wright}, \& {Fischer}}]{Cumming+08}
{Cumming}, A., {Butler}, R.~P., {Marcy}, G.~W., {Vogt}, S.~S., {Wright}, J.~T.,
  \& {Fischer}, D.~A. 2008, \pasp, 120, 531, 0803.3357

\bibitem[{{Eggleton} {et~al.}(1998){Eggleton}, {Kiseleva}, \& {Hut}}]{1998EKH}
{Eggleton}, P.~P., {Kiseleva}, L.~G., \& {Hut}, P. 1998, \apj, 499, 853,
  arXiv:astro-ph/9801246

\bibitem[{{Fabrycky} \& {Tremaine}(2007)}]{Dan}
{Fabrycky}, D., \& {Tremaine}, S. 2007, \apj, 669, 1298, 0705.4285

\bibitem[{{Fabrycky} \& {Winn}(2009)}]{FW09}
{Fabrycky}, D.~C., \& {Winn}, J.~N. 2009, \apj, 696, 1230, 0902.0737

\bibitem[{{Fang} \& {Margot}(2012)}]{Fang12}
{Fang}, J., \& {Margot}, J.-L. 2012, \apj, 761, 92, 1207.5250

\bibitem[{{H{\'e}brard} {et~al.}(2013){H{\'e}brard}, {Collier Cameron},
  {Brown}, {D{\'{\i}}az}, {Faedi}, {Smalley}, {Anderson}, {Armstrong},
  {Barros}, {Bento}, {Bouchy}, {Doyle}, {Enoch}, {G{\'o}mez Maqueo Chew},
  {H{\'e}brard}, {Hellier}, {Lendl}, {Lister}, {Maxted}, {McCormac}, {Moutou},
  {Pollacco}, {Queloz}, {Santerne}, {Skillen}, {Southworth}, {Tregloan-Reed},
  {Triaud}, {Udry}, {Vanhuysse}, {Watson}, {West}, \& {Wheatley}}]{Hebrard+13}
{H{\'e}brard}, G. {et~al.} 2013, \aap, 549, A134, 1211.0810

\bibitem[{{Hirano} {et~al.}(2012){Hirano}, {Narita}, {Sato}, {Takahashi},
  {Masuda}, {Takeda}, {Aoki}, {Tamura}, \& {Suto}}]{Hirano+12}
{Hirano}, T. {et~al.} 2012, \apjl, 759, L36, 1209.4362

\bibitem[{{Ho} \& {Turner}(2011)}]{Ho+11}
{Ho}, S., \& {Turner}, E.~L. 2011, \apj, 739, 26, 1003.4738

\bibitem[{{Huber} {et~al.}(2013){Huber}, {Carter}, {Barbieri}, {Miglio},
  {Deck}, {Fabrycky}, {Montet}, {Buchhave}, {Chaplin}, {Hekker},
  {Montalb{\'a}n}, {Sanchis-Ojeda}, {Basu}, {Bedding}, {Campante},
  {Christensen-Dalsgaard}, {Elsworth}, {Stello}, {Arentoft}, {Ford},
  {Gilliland}, {Handberg}, {Howard}, {Isaacson}, {Johnson}, {Karoff},
  {Kawaler}, {Kjeldsen}, {Latham}, {Lund}, {Lundkvist}, {Marcy}, {Metcalfe},
  {Silva Aguirre}, \& {Winn}}]{Huber+13}
{Huber}, D. {et~al.} 2013, Science, 342, 331, 1310.4503

\bibitem[{{Innanen} {et~al.}(1997){Innanen}, {Zheng}, {Mikkola}, \&
  {Valtonen}}]{Inn+97}
{Innanen}, K.~A., {Zheng}, J.~Q., {Mikkola}, S., \& {Valtonen}, M.~J. 1997,
  \aj, 113, 1915

\bibitem[{{Johnson} {et~al.}(2007){Johnson}, {Fischer}, {Marcy}, {Wright},
  {Driscoll}, {Butler}, {Hekker}, {Reffert}, \& {Vogt}}]{Johnson07}
{Johnson}, J.~A. {et~al.} 2007, \apj, 665, 785, 0704.2455

\bibitem[{{Kaib} {et~al.}(2011){Kaib}, {Raymond}, \& {Duncan}}]{Kaib11}
{Kaib}, N.~A., {Raymond}, S.~N., \& {Duncan}, M.~J. 2011, \apjl, 742, L24,
  1110.5911

\bibitem[{{Kratter} \& {Perets}(2012)}]{Kratter12}
{Kratter}, K.~M., \& {Perets}, H.~B. 2012, \apj, 753, 91, 1204.2014

\bibitem[{{Lai} {et~al.}(2010){Lai}, {Foucart}, \& {Lin}}]{Lai+10}
{Lai}, D., {Foucart}, F., \& {Lin}, D.~N.~C. 2010, ArXiv e-prints, 1008.3148

\bibitem[{{Li} \& {Loeb}(2013)}]{Li13}
{Li}, G., \& {Loeb}, A. 2013, \mnras, 429, 3040, 1209.1104

\bibitem[{{Li} {et~al.}(2014){Li}, {Naoz}, {Holman}, \& {Loeb}}]{Li+14}
{Li}, G., {Naoz}, S., {Holman}, M., \& {Loeb}, A. 2014, ArXiv e-prints,
  1405.0494

\bibitem[{{Li} {et~al.}(2013){Li}, {Naoz}, {Kocsis}, \& {Loeb}}]{Li+13}
{Li}, G., {Naoz}, S., {Kocsis}, B., \& {Loeb}, A. 2013, ArXiv e-prints,
  1310.6044

\bibitem[{{Lillo-Box} {et~al.}(2014){Lillo-Box}, {Barrado}, {Moya},
  {Montesinos}, {Montalb{\'a}n}, {Bayo}, {Barbieri}, {R{\'e}gulo}, {Mancini},
  {Bouy}, \& {Henning}}]{Lillo-Box14}
{Lillo-Box}, J. {et~al.} 2014, \aap, 562, A109, 1312.3943

\bibitem[{{Lin} \& {Papaloizou}(1986)}]{Lin+86}
{Lin}, D.~N.~C., \& {Papaloizou}, J. 1986, \apj, 309, 846

\bibitem[{{Lissauer} {et~al.}(2011){Lissauer}, {Ragozzine}, {Fabrycky},
  {Steffen}, {Ford}, {Jenkins}, {Shporer}, {Holman}, {Rowe}, {Quintana},
  {Batalha}, {Borucki}, {Bryson}, {Caldwell}, {Carter}, {Ciardi}, {Dunham},
  {Fortney}, {Gautier}, {Howell}, {Koch}, {Latham}, {Marcy}, {Morehead}, \&
  {Sasselov}}]{Lissauer+11}
{Lissauer}, J.~J. {et~al.} 2011, \apjs, 197, 8, 1102.0543

\bibitem[{{Mardling}(2010)}]{Mardling10}
{Mardling}, R.~A. 2010, \mnras, 407, 1048, 1001.4079

\bibitem[{{Masset} \& {Papaloizou}(2003)}]{Mass+03}
{Masset}, F.~S., \& {Papaloizou}, J.~C.~B. 2003, \apj, 588, 494,
  arXiv:astro-ph/0301171

\bibitem[{{Morton} \& {Johnson}(2011)}]{Morton10}
{Morton}, T.~D., \& {Johnson}, J.~A. 2011, \apj, 729, 138, 1010.4025

\bibitem[{{Moutou} {et~al.}(2011){Moutou}, {D{\'{\i}}az}, {Udry},
  {H{\'e}brard}, {Bouchy}, {Santerne}, {Ehrenreich}, {Arnold}, {Boisse},
  {Bonfils}, {Delfosse}, {Eggenberger}, {Forveille}, {Lagrange}, {Lovis},
  {Martinez}, {Pepe}, {Perrier}, {Queloz}, {Santos}, {S{\'e}gransan},
  {Toublanc}, {Troncin}, {Vanhuysse}, \& {Vidal-Madjar}}]{Moutou+11}
{Moutou}, C. {et~al.} 2011, \aap, 533, A113, 1105.3849

\bibitem[{{Murray} \& {Dermott}(2000)}]{MD00}
{Murray}, C.~D., \& {Dermott}, S.~F. 2000, {Solar System Dynamics}, ed.
  {Murray, C.~D.~\& Dermott, S.~F.}

\bibitem[{{Nagasawa} {et~al.}(2008){Nagasawa}, {Ida}, \& {Bessho}}]{Nag+08}
{Nagasawa}, M., {Ida}, S., \& {Bessho}, T. 2008, \apj, 678, 498, 0801.1368

\bibitem[{{Naoz} {et~al.}(2011){Naoz}, {Farr}, {Lithwick}, {Rasio}, \&
  {Teyssandier}}]{Naoz11}
{Naoz}, S., {Farr}, W.~M., {Lithwick}, Y., {Rasio}, F.~A., \& {Teyssandier}, J.
  2011, \nat, 473, 187, 1011.2501

\bibitem[{{Naoz} {et~al.}(2013){Naoz}, {Farr}, {Lithwick}, {Rasio}, \&
  {Teyssandier}}]{Naoz13}
------. 2013, \mnras, 431, 2155, 1107.2414

\bibitem[{{Naoz} {et~al.}(2012){Naoz}, {Farr}, \& {Rasio}}]{Naoz+12bin}
{Naoz}, S., {Farr}, W.~M., \& {Rasio}, F.~A. 2012, \apjl, 754, L36, 1206.3529

\bibitem[{{Nordhaus} {et~al.}(2010){Nordhaus}, {Spiegel}, {Ibgui}, {Goodman},
  \& {Burrows}}]{Nordhaus10}
{Nordhaus}, J., {Spiegel}, D.~S., {Ibgui}, L., {Goodman}, J., \& {Burrows}, A.
  2010, \mnras, 408, 631, 1002.2216

\bibitem[{{Paczy{\'n}ski}(1971)}]{Paczyski71}
{Paczy{\'n}ski}, B. 1971, \araa, 9, 183

\bibitem[{{Paxton} {et~al.}(2011){Paxton}, {Bildsten}, {Dotter}, {Herwig},
  {Lesaffre}, \& {Timmes}}]{Paxton+11}
{Paxton}, B., {Bildsten}, L., {Dotter}, A., {Herwig}, F., {Lesaffre}, P., \&
  {Timmes}, F. 2011, \apjs, 192, 3, 1009.1622

\bibitem[{{Paxton} {et~al.}(2013){Paxton}, {Cantiello}, {Arras}, {Bildsten},
  {Brown}, {Dotter}, {Mankovich}, {Montgomery}, {Stello}, {Timmes}, \&
  {Townsend}}]{Paxton+13}
{Paxton}, B. {et~al.} 2013, \apjs, 208, 4, 1301.0319

\bibitem[{{Rasio} \& {Ford}(1996)}]{RF96}
{Rasio}, F.~A., \& {Ford}, E.~B. 1996, Science, 274, 954

\bibitem[{{Rogers} {et~al.}(2012){Rogers}, {Lin}, \& {Lau}}]{Rogers+12}
{Rogers}, T.~M., {Lin}, D.~N.~C., \& {Lau}, H.~H.~B. 2012, \apjl, 758, L6,
  1209.2435

\bibitem[{{Rogers} {et~al.}(2013){Rogers}, {Lin}, {McElwaine}, \&
  {Lau}}]{Rogers+13IGW}
{Rogers}, T.~M., {Lin}, D.~N.~C., {McElwaine}, J.~N., \& {Lau}, H.~H.~B. 2013,
  \apj, 772, 21, 1306.3262

\bibitem[{{Sanchis-Ojeda} {et~al.}(2012){Sanchis-Ojeda}, {Fabrycky}, {Winn},
  {Barclay}, {Clarke}, {Ford}, {Fortney}, {Geary}, {Holman}, {Howard},
  {Jenkins}, {Koch}, {Lissauer}, {Marcy}, {Mullally}, {Ragozzine}, {Seader},
  {Still}, \& {Thompson}}]{Sanchis+12}
{Sanchis-Ojeda}, R. {et~al.} 2012, \nat, 487, 449, 1207.5804

\bibitem[{{Sato} {et~al.}(2008){Sato}, {Toyota}, {Omiya}, {Izumiura}, {Kambe},
  {Masuda}, {Takeda}, {Itoh}, {Ando}, {Yoshida}, {Kokubo}, \& {Ida}}]{Sato08}
{Sato}, B. {et~al.} 2008, \pasj, 60, 1317, 0807.0268

\bibitem[{{Schlaufman} \& {Winn}(2013)}]{Schlaufman13}
{Schlaufman}, K.~C., \& {Winn}, J.~N. 2013, \apj, 772, 143, 1306.0567

\bibitem[{{Spalding} \& {Batygin}(2014)}]{Spalding14}
{Spalding}, C., \& {Batygin}, K. 2014, \apj

\bibitem[{{Steffen} \& {Farr}(2013)}]{Steffen+13}
{Steffen}, J.~H., \& {Farr}, W.~M. 2013, \apjl, 774, L12, 1306.3526

\bibitem[{{Swift} {et~al.}(2013){Swift}, {Johnson}, {Morton}, {Crepp},
  {Montet}, {Fabrycky}, \& {Muirhead}}]{Swift+13}
{Swift}, J.~J., {Johnson}, J.~A., {Morton}, T.~D., {Crepp}, J.~R., {Montet},
  B.~T., {Fabrycky}, D.~C., \& {Muirhead}, P.~S. 2013, \apj, 764, 105,
  1301.0023

\bibitem[{{Takeda} {et~al.}(2008){Takeda}, {Kita}, \& {Rasio}}]{Takeda}
{Takeda}, G., {Kita}, R., \& {Rasio}, F.~A. 2008, \apj, 683, 1063, 0802.4088

\bibitem[{{Tamayo} {et~al.}(2013){Tamayo}, {Burns}, {Hamilton}, \&
  {Nicholson}}]{Tamayo+13}
{Tamayo}, D., {Burns}, J.~A., {Hamilton}, D.~P., \& {Nicholson}, P.~D. 2013,
  \aj, 145, 54, 1212.0028

\bibitem[{{Teitler} \& {K{\"o}nigl}(2014)}]{Teitler+14}
{Teitler}, S., \& {K{\"o}nigl}, A. 2014, ArXiv e-prints, 1403.5860

\bibitem[{{Thies} {et~al.}(2011){Thies}, {Kroupa}, {Goodwin}, {Stamatellos}, \&
  {Whitworth}}]{Thies+11}
{Thies}, I., {Kroupa}, P., {Goodwin}, S.~P., {Stamatellos}, D., \& {Whitworth},
  A.~P. 2011, \mnras, 417, 1817, 1107.2113

\bibitem[{{Tremaine}(2011)}]{Tremaine11}
{Tremaine}, S. 2011, in IAC Talks, Astronomy and Astrophysics Seminars from the
  Instituto de Astrof{\'{\i}}sica de Canarias, 227

\bibitem[{{Tremaine} {et~al.}(2009){Tremaine}, {Touma}, \&
  {Namouni}}]{Tremaine+09}
{Tremaine}, S., {Touma}, J., \& {Namouni}, F. 2009, \aj, 137, 3706, 0809.0237

\bibitem[{{Triaud} {et~al.}(2010){Triaud}, {Collier Cameron}, {Queloz},
  {Anderson}, {Gillon}, {Hebb}, {Hellier}, {Loeillet}, {Maxted}, {Mayor},
  {Pepe}, {Pollacco}, {S{\'e}gransan}, {Smalley}, {Udry}, {West}, \&
  {Wheatley}}]{Tri+10}
{Triaud}, A.~H.~M.~J. {et~al.} 2010, \aap, 524, A25+, 1008.2353

\bibitem[{{Trilling} {et~al.}(1998){Trilling}, {Benz}, {Guillot}, {Lunine},
  {Hubbard}, \& {Burrows}}]{TrillingBGLHB1998}
{Trilling}, D.~E., {Benz}, W., {Guillot}, T., {Lunine}, J.~I., {Hubbard},
  W.~B., \& {Burrows}, A. 1998, \apj, 500, 428, arXiv:astro-ph/9801292

\bibitem[{{Valsecchi} {et~al.}(2012){Valsecchi}, {Farr}, {Willems}, {Deloye},
  \& {Kalogera}}]{Valsecchi+12}
{Valsecchi}, F., {Farr}, W.~M., {Willems}, B., {Deloye}, C.~J., \& {Kalogera},
  V. 2012, \apj, 745, 137, 1105.4837

\bibitem[{{Valsecchi} \& {Rasio}(2014{\natexlab{a}})}]{ValsecchiRasio14b}
{Valsecchi}, F., \& {Rasio}, F.~A. 2014{\natexlab{a}}, ArXiv e-prints,
  1403.1870

\bibitem[{{Valsecchi} \& {Rasio}(2014{\natexlab{b}})}]{ValsecchiRasio14a}
------. 2014{\natexlab{b}}, ArXiv e-prints, 1402.3857

\bibitem[{{Van Eylen} {et~al.}(2014){Van Eylen}, {Lund}, {Silva Aguirre},
  {Arentoft}, {Kjeldsen}, {Albrecht}, {Chaplin}, {Isaacson}, {Pedersen},
  {Jessen-Hansen}, {Tingley}, {Christensen-Dalsgaard}, {Aerts}, {Campante}, \&
  {Bryson}}]{Van14}
{Van Eylen}, V. {et~al.} 2014, \apj, 782, 14, 1312.4938

\bibitem[{{Veras} {et~al.}(2013){Veras}, {Hadjidemetriou}, \& {Tout}}]{Veras13}
{Veras}, D., {Hadjidemetriou}, J.~D., \& {Tout}, C.~A. 2013, \mnras, 435, 2416,
  1308.0599

\bibitem[{{Winn} {et~al.}(2010){Winn}, {Fabrycky}, {Albrecht}, \&
  {Johnson}}]{Winn+10b}
{Winn}, J.~N., {Fabrycky}, D., {Albrecht}, S., \& {Johnson}, J.~A. 2010, \apjl,
  718, L145, 1006.4161

\bibitem[{{Wu} \& {Lithwick}(2011)}]{WY11}
{Wu}, Y., \& {Lithwick}, Y. 2011, \apj, 735, 109, 1012.3475

\end{thebibliography}

\end{document}